# Comprehensive Security Framework for Global Threads Analysis

Jacques SARAYDARYAN, Fatiha BENALI and Stéphane UBEDA

[1] Exaprotect R&D
Villeurbanne, 69100, France
*jsaraydaryan@exaprotect.com*

[2] INSA Lyon
Villeurbanne, 69100, France
*Fatiha.benali@insa-lyon.fr*

[3] INSA Lyon
Villeurbanne, 69100, France
*Stephane.ubeda@insa-lyon.fr*

**Abstract**

Cyber criminality activities are changing and becoming more and more professional. With the growth of financial flows through the Internet and the Information System (IS), new kinds of thread arise involving complex scenarios spread within multiple IS components. The IS information modeling and Behavioral Analysis are becoming new solutions to normalize the IS information and counter these new threads. This paper presents a framework which details the principal and necessary steps for monitoring an IS. We present the architecture of the framework, i.e. an ontology of activities carried out within an IS to model security information and User Behavioral analysis. The results of the performed experiments on real data show that the modeling is effective to reduce the amount of events by 91%. The User Behavioral Analysis on uniform modeled data is also effective, detecting more than 80% of legitimate actions of attack scenarios.

## 1. Introduction

Today, information technology and networking resources are dispersed across an organization. Threats are similarly distributed across many organization resources. Therefore, the Security of information systems (IS) is becoming an important part of business processes. Companies must deal with open systems on the one hand and ensure a high protection on the other hand. As a common task, an administrator starts with the identification of threats related to business assets, and applies a security product on each asset to protect an IS. Then, administrators tend to combine and multiply security products and protection techniques such as firewalls, antivirus, Virtual Private Network (VPN), Intrusion Detection System (IDS) and security audits.

But are the actions carried out an IS only associated with attackers? Although the real figures are difficult to know, most experts agree that the greatest threat for security comes not only from outside, but also from inside the company. Now, administrators are facing new requirements consisting in tracing the legitimate users. Do we need to trace other users of IS even if they are legitimate? Monitoring attackers and legitimate users aims at detecting and identifying a malicious use of the IS, stopping attacks in progress and isolating the attacks that may occur, minimizing risks and preventing future attacks to take counter measures. To trace legitimate users, some administrators perform audit on applications, operating systems and administrators products. Events triggered by these mechanisms are thus relevant for actions to be performed by legitimate users on these particular resources.

Monitoring organization resources produces a great amount of security-relevant information. Devices such as firewalls, VPN, IDS, operating systems and switches may generate tens of thousands of events per second. Security administrators are facing the task of analyzing an increasing number of alerts and events. The approaches implemented in security products are different, security products analysis may not be exact, they may produce false positives (normal events considered as attacks) and false negatives (Malicious events considered as normal). Alerts and events can be of different natures and level of granularity; in the form of logs, Syslog, SNMP traps, security alerts and other reporting mechanisms. This information is extremely valuable and the operations that must be carried out on security require a constant analysis of these data to guarantee knowledge on threats in real time. An appropriate treatment for these issues is not trivial and needs a large range of knowledge. Until recently, the combined security status of an organization could not be decided. To compensate for this failure, attention must be given to integrate local security disparate observations into a single view of the composite security state of an organization.

To address this problem, both vendors and researchers have proposed various approaches. Vendors' approaches are referred to as Security Information Management (SIM) or Security Event Management (SEM). They address a company's need to manage alerts, logs and events, and any other security elementary information coming from company resources such as networking devices of all sorts, diverse security products (such as firewalls, IDS and antivirus), operating systems, applications and databases. The purpose is to create a good position for observation from which an enterprise can manage threats, exposure, risk, and vulnerabilities. The industry' approaches focus on information technology events in addition to security event. They can trace IS user, although the user is an attacker or a legitimate user. The intrusion detection research community has developed a number of different approaches to make security products interact. They focus on the correlation aspect in the analysis step of data, they do not provide insights into what properties of the data being analyzed.

The question asked in this article is to know what is missing in today's distributed intrusion detection. However, it is not clear how the different parts that compose Vendor product should be. Vendor's approaches do not give information on how data are modeled and analyzed. Moreover, vendors claim that they can detect attacks, but how can they do if the information is heterogeneous? How can they rebuild IS misuse scenarios? All the same, research works lack of details on the different components, which make the correlation process effective. They were developed in particular environments. They rarely address the nature of the data to be analyzed, they do not give global vision of the security state of an IS because some steps are missing to build the IS scenarios of use. Both approaches do not indicate how they should be implemented and evaluated. Therefore, a coherent architecture and explanation of a framework, which manages company's security effectively is needed.

The framework must collect and normalize data across a company structure, then cleverly analyze data in order to give administrators a global view of the security status within the company. It can send alerts to administrators so that actions can be taken or it can automate responses that risks can be addressed and remediated quickly, by taking actions such as shutting down an account of a legitimate user who misuses the IS or ports on firewalls.

The distributed architecture concept, DIDS (Distributive Intrusion Detection System), first appeared in 1989 (Haystack Lab). This first analysis of distributed information did not present a particular architecture but collected the information of several audit files on IS hosts. The recent global IS monitoring brings new challenges in the collection and analysis of distributed data. Recent distributed architectures are mostly based on Agents. These types of architectures are mainly used in research projects and commercial solutions (Arcsight, Netforensic, Intellitactics, LogLogic). An agent is an autonomy application with predefined goals [31]. These goals are various: monitor an environment, deploy counter-measures, pre-analyze information, etc. The autonomy and goal of an agent would depend on a used architecture. Two types of architecture can be highlighted, distributive centralized architecture and distributive collaborative architecture. Zheng Zhang et al. [1] provided a hierarchical centralized architecture for network attacks detection. The authors recommend a three-layer architecture which collects and analyzes information from IS components and from other layers. This architecture provides multiple levels of analysis for the network attacks detection; a local attack detection provided by the first layer and a global attack detection provided by upper layers. A similar architecture was provided by [39] for the network activity graph construction revealing local and global casual structures of the network activity. K. Boudaoud [4] provides a hierarchical collaborative architecture. Two main layers are used. The first one is composed of agents which analyze local components to discover intrusion based on their analysis of their own knowledge but also with the knowledge of other agents. The upper layer collects information from the first layer and tries to detect global attacks. In order to detect intrusions, each agent holds attacks signatures (simple pattern for the first layer, attack graph for the second layer). Helmer et al. [13] provide a different point of view by using mobile agents. A light weight agent has the ability to "travel" on different data sources. Each mobile agent uses a specific schema of analysis (Login Failed, System Call, TCP connection) and can communicate with other agents to refine their analyses.

Despite many discussions, scalability, analysis availability and collaborative architecture are difficult to apply, in today's, infrastructure but also time and effort consuming.

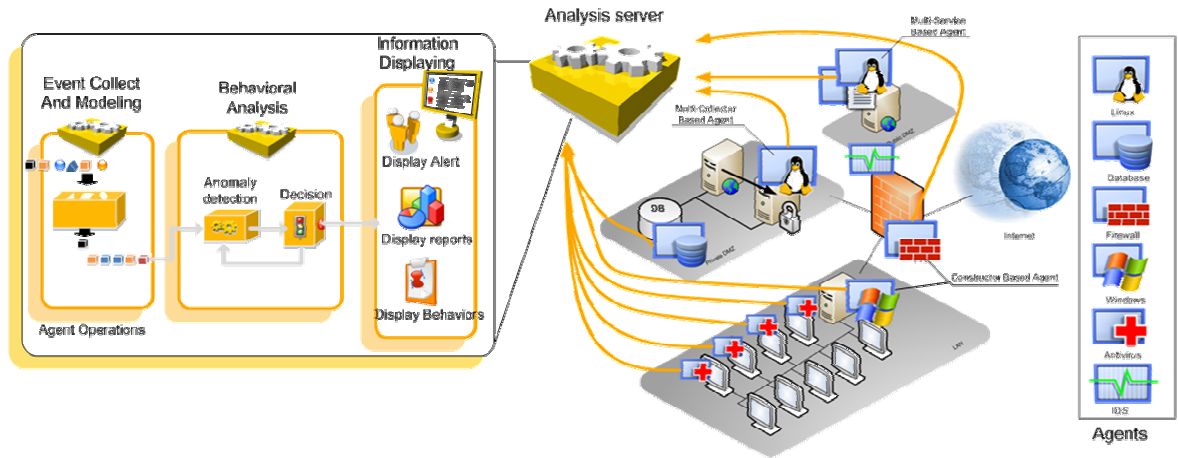

Fig. 1 Global Anomaly Intrusion Detection Architecture

Thus, despite known drawbacks, distributive centralized architectures will be used in our approach for the analysis of distributive knowledge in the IS.

All IS and User behaviors' actions are distributed inside IS components. In order to collect and analyze these knowledge, we propose an architecture composed of distributed agents allowing distributive data operations. Distributive agent aims at collecting data by making pre-operations and forwarding this information to an Analysis Server. The Analysis Server holds necessary information to correlate and detect abnormal IS behaviors. This architecture is a hierarchical central architecture. Distributive agents share two main functionalities:

- a collector function aiming at collecting information on monitored components,
- an homogenization function aiming at standardizing and filtering collected information.

As shown in figure 1, three types of agents are used. The constructor-based agent aims at collecting information from a specific IS components (Window Host, Juniper firewall).

The multi-collector based agent aims at collecting information from several IS components redirecting their flow of log (syslog). Then, the multi-service based agent aims at collecting several different information (system log, Web server application log) from a single IS component.

This paper presents a comprehensive framework to manage information security intelligently so that processes implemented in analysis module are effective. We focus our study on the information modeling function, the information volume reductions and the Abnormal Users Behavior detection. A large amount of data triggered in a business context is then analyzed by the framework. The results show that the effectiveness of the analysis process is highly dependent on the data modeling, and that unknown attack scenarios could be efficiently detected without hard pre-descriptive information. Our decision module also allows reducing false positive.

The reminder of this paper is structured as follows. In the next section, related work on security event modeling and behavioral analysis is covered. In the third section, the proposed modeling for event security in the context of IS global vision is presented. Section 4 details the anomaly detection module. The validation of the homogenization function and the anomaly detection module is performed on real data and presented in Section 5. Finally, the conclusions and perspectives of our work are mentioned in the last section.

## 2. Related Work

As mentioned in the introduction, security monitoring of an IS is strongly related to the information generated in products' log file and to the analysis carried out on this information. In this section, we address both event modeling and Behavioral Analysis state of the art.

### 2.1 Event Modeling

All the research works performed on information security modeling direct our attention on describing attacks. There is a lack of describing information security in the context of a global vision of the IS security introduced in the previous section. As events are generated in our framework by different products, events can be represented in different formats with a different vocabulary. Information modeling aims to represent each product event into a common format. The common format requires a common specification of the semantics and the syntax of the events. There is a high number of alerts classification proposed for use in intrusion detection research. Four approaches were used to describe attacks: list of terms, taxonomies, ontologies and attacks language. The easiest classification proposes a list of single terms [7, 18], covering various aspects of attacks. The number of terms differs from an author to another one. Other authors have created categories regrouping many terms under a common definition. Cheswick and Bellovin classify attacks into seven categories [5]. Stallings classification [38] is based on the action. The model focuses on transiting data and defines four categories of attacks: interruption, interception, modification and fabrication. Cohen [6] groups attacks into categories that describe the result of an attack. Other authors developed categories based on empirical data. Each author uses an events corpus generated in a specific environment. Neumann and Parker [25] works were based on a corpus of 3000 incidents collected for 20 years; they created nine classes according to attacking techniques. Terms tend to not be mutually exclusive; this type of classification can not provide a classification scheme that avoids ambiguity.

To avoid these drawbacks, a lot of taxonomies were developed to describe attacks. Neumann [24] extended the classification in [25] by adding the exploited vulnerabilities and the impact of the attack. Lindqvist and Jonson [21] presented a classification based on the Neumann classification [25]. They proposed intrusion results and intrusion techniques as dimension for classification. John Howard [16] presented a taxonomy of computer and network attacks. The taxonomy consists in five dimensions: attackers, tools, access, results and objectives. The author worked on the incidents of the Computer Emergency Response Team (CERT), the taxonomy is a process-driven. Howard extends his work by refining some of the dimensions [15]. Representing attacks by taxonomies is an improvement compared with the list of terms: individual attacks are described with an enriched semantics, but taxonomies fail to meet mutual exclusion requirements, some of the categories may overlap. However, the ambiguity problem still exists with the refined taxonomy.

Undercoffer and al [3] describe attacks by an ontology. Authors have proposed a new way of sharing the knowledge about intrusions in distributed IDS environment. Initially, they developed a taxonomy defined by the target, means, consequences of an attack and the attacker. The taxonomy was extended to an ontology, by defining the various classes, their attributes and their relations based on an examination of 4000 alerts. The authors have built correlation decisions based on the knowledge that exists in the modeling. The developed ontology represents the data model for the triggered information by IDSs.

Attack languages are proposed by several authors to detect intrusions. These languages are used to describe the presence of attacks in a suitable format. These languages are classified in six distinct categories presented in [12]: Exploit languages, event languages, detection languages, correlation languages, reporting languages and response languages. The Correlation languages are currently the interest of several researchers in the intrusion detection community. They specify relations between attacks to identify numerous attacks against the system. These languages have different characteristics but are suitable for intrusion detection, in particular environments. Language models are based on the models that are used for describing alerts or events semantic. They do not model the semantics of events but they implicitly use taxonomies of attacks in their modeling.

All the researches quoted above only give a partial vision of the monitored system, they were focused on the conceptualization of attacks or incidents, which is due to the consideration of a single type of monitoring product which is the IDS. It is important to mention the efforts done to realize a data model for information security. The first attempts were undertaken by the American agency - Defense Advanced Research Projects Agency (DARPA), which has created the Common Intrusion Detection Framework (CIDF) [32]. The objective of the CIDF is to develop protocols and applications so that intrusion detection research projects can share information. Work on CIDF was stopped in 1999 and this format was not implemented by any product. Some ideas introduced in the CIDF have encouraged the creation of a work group called Intrusion Detection Working Group (IDWG) at Internet Engineering Task Force (IETF). IETF have proposed the Intrusion Detection Message Exchange Format (IDMEF) [8] as a way to set a standard representation for intrusion alerts. IDMEF became a standard format with the RFC 4765[1]. The effort of the IDMEF is centered on alert syntax representation. In the implementations of IDSs, each IDS chooses the name of the attack, different IDSs can give different names to the same attack. As a result, similar information can be tagged differently and handled as two different alerts.

Modeling information security is a necessary and important task. Information security is the input data for all the analysis processes, e.g. the correlation process. All the analysis processes require automatic processing of information. Considering the number of alerts or events generated in a monitored system, the process, which manages this information, must be able to think on these data. We need an information security modeling based on abstraction of deployed products and mechanisms, which helps the classification process, avoids ambiguity to classify an event, and reflects the reality. Authors in [16,21] agree that the proposed classification for intrusion detection must have the following characteristics: accepted, unambiguous, understandable, determinist, mutually exclusive, exhaustive. To ensure the presence of all these characteristics, it is necessary to use an ontology to describe the semantics of security information.

---

[1] http://www.rfc-editor.org/rfc/rfc4765.txt

2.2 Behavioral Analysis

Even if Host Intrusion Detection System (HIDS) and Network Intrusion Detection System (NIDS) tools are known to be efficient for local vision by detecting or blocking unusual and forbidden activities, they can not detect new attack scenarios involving several network components. Focusing on this issue, industrial and research communities show a great interest in the Global Information System Monitoring.

Recent literatures in the intrusion detection field [30] aim at discovering and modeling global attack scenarios and Information System dependencies (IS components relationships). In fact, recent approaches deal with the Global Information System Monitoring like [22] who describes a hierarchical attack scenario representation. The authors provide an evaluation of the most credible attacker's step inside a multistage attack scenario. [28] computes also attack scenario graphs through the association of vulnerabilities on IS components and determines a "distance" between correlated events and these attack graphs. In the same way, [26] used a semi-explicit correlation method to automatically build attack scenarios. With a pre-processing stage, the authors model pre-conditions and post conditions for each event. The association of pre and post conditions of each event leads to the construction of graphs representing attack scenarios. Other approaches automatically discover an attack scenario with model checking methods, which involves a full IS component interaction and configuration description [36].

However, classical intrusion detection schemes are composed of two types of detection: Signature based and Anomaly based detections. The anomaly detection is not developed regarding to Global IS Monitoring. Few approaches intend to model system normal behavior. Authors in [11] model IS components' interactions in order to discover causes of IS disaster (Forensic Analysis). The main purpose of this approach is to build casual relationships between IS components to discover the origin of an observed effect. The lack of anomaly detection System can be explained by the fact that working on the Global vision introduces three main limitations. First of all, the volume of computed data can reach thousands of events per second. Secondly, collected information is heterogeneous due to the fact that each IS component holds its own events description. Finally, the complexity of attacks scenarios and IS dependencies increases very quickly with the volume of data.

## 3. Event Modeling

As we previously stated, managing information security has to deal with the several differences existing in the monitoring products. To achieve this goal, it is necessary to transform raw messages in a uniform representation. Indeed, all the events and alerts must be based on the same semantics description, and be transformed in the same data model. To have a uniform representation of semantics, we focus on concepts handled by the products, we use them to describe the semantics messages. In this way, we are able to offset products types, functions, and products languages aside. The Abstraction concept was already evoked in intrusions detection field by Ning and Al [27]. Authors consider that the abstraction is important for two primary reasons. First, the systems to be protected as well as IDSs are heterogeneous. In particular, a distributed system is often composed of various types of heterogeneous components. Abstraction becomes thus a necessary means to hide the difference between these component systems, and to allow the detection of intrusions in the distributed systems. Secondly, abstraction is often used to remove all the non relevant details, so that IDS can avoid an useless complexity and concentrate on the essential information.
The description of the information generated by a deployed solution is strongly related to the action perceived by the system, this action can be observed at any time of its life cycle: its launching, its interruption or its end. An event can inform that: an action has just started, it is in progress, it failed or it is finished. To simplify, we retained information semantics modeling via the concept of observed action. We obtain thus a modeling that fits to any type of observation, and meets the abstraction criteria.

3.1 Action Theory
In order to model the observed action, we refer to the works that have already been done in the Action Theory of the philosophy field. According to the traditional model of the action explained by the authors in [9,19], an action is an Intention directed to an Object and uses a Movement. It is generally conceded that intentions have a motivation role for the action. Authors underline that this role is not limited in starting the action but in supporting it until its completion. Our actions utilize movements, explanation of the action remains incomplete if we do not take into account the movements. Movement is the means to achieve the action. The object is the target towards which the action is directed to. In summary, the human actions are in conformity with a certain logic. To say that an agent carries out an action A, it is to say that the agent had an Intention I, by making a Movement M to produce an effect on a Target T of Action A. Action's basic model is so composed by the following concepts: intention, movement and target.

### 3.2 Event Semantics

We observe the action performed in the monitored system and see that this action is dissociating from the human mind. We add another concept, i.e. the Effect, to the basic model (Intention, Movement and Target) of the Action Theory. We can say that this modeling is a general modeling, it can be adapted to any context of the monitoring such as in IS intruders monitoring, or in the monitoring of bank physical intruders. All we have to do, is to instantiate the meta-model with the intrusion detection context's vocabulary.

We have outlined an adaptation of this meta-model to our context of the IS monitoring from threats. The concepts are redefined as follows:
- **Intention:** the objective for which the user carries out his action,
- **Movement:** the means used to carry out the objective of the user,
- **Target:** the resource in the IS to which the action is directed to,
- **Gain:** the effect produced by the action on the system, i.e. if the user makes a success of his attempt to carry out an action or not.

Security information is an Intention directed towards a Target which uses a Movement to reach the target, and produces a Gain. Figure 2 illustrates the ontology concepts, the semantic relation between the concepts, and the general semantics of a security event.

In order to identify the intentions of a user's IS, we have studied the attacker strategy [17,26]. In fact, once the attacker is in the IS, he can undergo both attacker's and legitimate user's action.  Analyzing attacker strategy provides an opportunity to reconstruct the various steps of an attack scenario or an IS utilization scenario, and perform pro-active actions to stop IS misuse. According to the attacker strategy, we have identified four intentions:

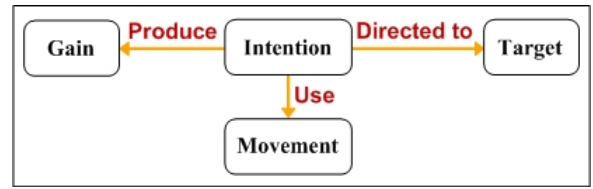

Fig. 2 Security message semantics

- **Recon:** intention of collecting information on a target,
- **Authentication:** intention to access to the IS via an authentication system,
- **Authorization:** intention to access to a resource of an IS,
- **System:** intention of modifying the availability of an IS resources.

Intentions are carried out through movements. We have categorized the movement into seven natures of movements:
- **Activity:** all the movements related to activities which do not change the configuration of the IS,
- **Config:** all the movements which change configuration,
- **Attack:** all the movements related to attacks,
- **Malware:** all the movements related to malwares. Malware are malicious software programs designed specifically to damage or disrupt a system,
- **Suspicious:** all the movements related to the suspicious activities detected by the products. In some cases, a product generates an event to inform that there was a suspicious activity on a resource, this information is reported by our modeling as it is.
- **Vulnerability:** all the movements related to vulnerabilities.
- **Information:** the probes can produce messages which do not reflect the presence of the action, they reflect a state observed on the IS.

Under each one of these main modes, we have identified the natures of movements. A movement is defined by a mode of movements (such as Activity, Config, Information, Attack, Suspicious, Vulnerability or Malware) and a nature of movement (such as Login, Read, Execute, etc.), the mode and the nature of the action defines the movement. As the model must be adapted to the context of the global vision of IS's security monitoring, it is clear that we have defined movements of presumed normal actions or a movement of presumed dangerous actions. An intention is able to have several movements, for example, an access to the IS performed by the opening of a user's session or by an account's configuration or by a bruteforce attack.

Each IS resource can be a target of an IS user activity. Targets are defined according to intentions.

- In the case of the Recon intention, an activity of information collection is carried out on a host. The target represents the host on whom the activity was detected.
- In the case of the Authentication intention, an activity of access to an IS is always carried out under the control of an authentication service. The target is represented by a pair (target1, target2), target1 and target2 refer respectively to the system that performs the authentication process and the object or an attribute of the object that authenticates on the authentication service.
- In the case of the Authorization intention, an access to an IS resource is always carried out under the control of a service, this service allows the access to a resource based on access criteria. The target is represented by a pair (target1, target2). Target1 and Target2 refer respectively to the resource which filters the accesses (which manages the rights of the users or groups) and the resource on which rights were used to reach it.
- In the case of the System intention, an activity depends on the availability of the system. The target is represented by a pair (target1, target2). Target1 and Target2 refer respectively to a resource and a property of the resource, for example (Host, CPU).

These constraints on the targets enable us to fix the level of details to be respected in the modeling. We have defined the Gain in the IS according to the Movement mode.

- In the case of the Activity and Config movement mode, Gain takes the values: Success, Failed, Denied or Error.
- In the case of the Malware, Vulnerability, Suspicious and Attack movement mode, Gain takes the value Detected.
- In the case of the Information movement mode, the event focuses on information about the system state. Gain is related to information on control and takes the values: Valid, Invalid or Notify, or related to information on thresholds and takes the values Expired, Exceeded, Low or Normal.

The result is an ontology described by four concepts: Intention, Movement, Target, Gain, tree semantics relations: Produce, Directed to, Use between the concepts, a vocabulary adapted to the context of the IS monitoring against security violation, and rules explaining how to avoid the ambiguity. For example, for an administrator action who succeeded in opening a session on a firewall, the ontology describes this action by the 4-uplets: Authentication (refers to the intention of the user), Activity Login (refers to the movement carried out by the user), Firewall Admin (refers to the target of the action carried out) and Success (refers to the result of the action). The category of the message to which this action belongs is:

Authentication_Activity.Login_Firewall.Admin_Successes.
We have identified 4 Intentions, 7 modes of Movement, 52 natures of Movement, 70 Targets and 13 Gains.

3.3 Event Data Model

It seems reasonable that the data model for information security can be based on standards. We have mentioned in 2.1 that the format IDMEF becomes a standard. We use this data model like a data model for event generated in a products interoperability context.

This format is composed of two subclasses Alert and Heartbeat. When the analyzer of an IDS detects an event for which he has been configured, it sends a message to inform their manager. Alert class is composed of nine classes: Analyzer, CreateTime, DetectTime, analyserTime, Source, Target, Classification, Assessment, and AdditionalData.

The IDMEF format Classification class is considered as a way to describe the alert semantics. The ontology developed in this framework describes all the categories of activities that can be undertaken in an IS. We define the Classification of the IDMEF data model class by a category of the ontology that reflects the semantics of the triggered raw event. Figure 3 illustrates the format IDMEF with modification of the class Classification.
Finally, with the proposed ontology and the adapted IDMEF data
model to information security in the context of global IS view, information is homogeneous. Indeed, all processes that can be implemented in the analysis server can be undertaken including the behavioral analysis.

## 4. Behavioral Analysis

Anomaly Detection System differs from signature based Intrusion Detection System by modeling normal reference instead of detecting well known patterns. Two periods are distinguished in Anomaly Detection: a first period, called training period, which builds and calibrates the normal reference. The detection of deviant events is performed during a second period called exploitation. We propose an Anomaly Detection System composed of four main blocks as shown in figure 4. The Event Collection and Modeling block aims at collecting normalized information from the different agents. The Event Selection block would filter only relevant information (see section 4.3). The Anomaly Detection block would model user's behaviors through an activity graph and a Bayesian Network (see section 4.1.1) during a training period and would detect anomaly (see section 4.2) in the exploitation period. Then all behavioral anomalies are evaluated by the Anomaly Evaluation block which identifies normal reference update from dangerous behavioral anomalies (see section 4.4). Each block will be explained in the following sections.

4.1 Model Method Selection

Modeling user behavior is a well-known topic in NIDS or HIDS regarding local component monitoring. The proliferation of security and management equipment units over the IS brings a class of anomaly detection. Following user behaviors over the IS implies new goals that anomaly detection needs to cover.

First of all we need to identify the owner of each event occurring on the IS. Then our model should be able to hold attributes identifying this owner. The fact that all users travel inside the IS implies that the user activity model should models sequences of events representing each user's actions on all IS components. Moreover, user behavior can be assimilated to a dynamic system. Modeling user activities should enhance periodical phenomena and isolate sporadic ones. Then, user behaviors hold major information of the usage of the system and can highlight the users' behaviors compliance

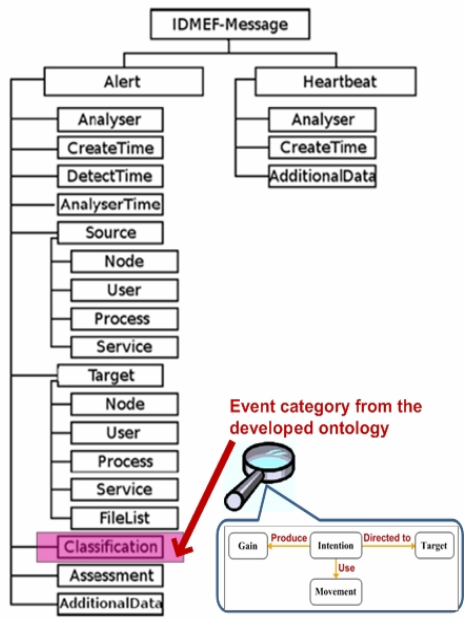

Fig. 3 The IDMEF data model with the class Classification represented by the category of the ontology that describes the with the security policy. This property should offer the ability to make the model fit the IS policies or plan future system evolutions to a security analyst. To achieve that, user activities modeling should be Human readable.

Several modeling methods are used for normal reference modeling in Anomaly Detection (e.g. classification, Neural Network, States Automaton, etc). Nevertheless three of them deal with the new goals: Hidden Markov Model, stochastic Petri Network and Bayesian Network. In Hidden Markov Model and stochastic Petri Network methods each node of sequences identifies one unique system or event state. Modeling the events of each user on each IS components would lead to the construction of a huge events graph. All these techniques can model probabilistic sequences of events but only Bayesian Network provides a human understandable model.

Bayesian Networks (BN) are well adapted for user's activities modeling regarding the Global Monitoring goals and provide a suitable model support. BN is a probabilistic graphical model that represents a set of variables and their conditional probabilities. BNs are built around an oriented acyclic graph which represents the structure of the network. This graph describes casual relationships between the variables. By instantiating a variable, each conditional probability is computed using mechanism of inference and the BN gives us the probabilities of all variables regarding this setting. By associating each node to a variable and each state of a node to a specific value, BN graph contracts knowledge in human readable graph. Furthermore, BNs are useful for learning probabilities in pre-computed data set and are well appropriate for the deviance detection. BN inference allows injecting variable values in BN graph and determining all conditional probabilities related to the injected proof.

To achieve a user activity Bayesian Network model, we need to create a Bayesian Network structure. This structure would refer to a graph of events where each node represents an event and each arc a causal relationship between two events. Some approaches used learning methods (k2 algorithm [11]) to reveal a Bayesian structure in a dataset. In the context of a global events monitoring, lots of parameters are involved and without some priori knowledge, self learning methods would extract inconsistent relationships between events.

Following our previous work [34] on user behaviors analysis, we specify three event's attributes involved in the identification of casual relationships: user login, IP address Source and Destination. Here, we enhance the fact that legitimate users performed two types of actions: local actions and remote actions. First, we focus our attention on event's attributes identifying local user action. The couple of attributes 'Source IP address' and 'user name' is usually used to identify users. These two attributes allow tracking user activities in a same location (e.g. work station, application server). To monitor remote user actions only, 'Destination IP address' and 'source IP address' attributes can be used. Then, to monitor a physical user location move, only the 'user login name' can be used to follow them.

These three attributes, i.e. user login, IP address Source and destination, are combined to determine correlation rules between two collected events as defined in our work [33]. Two events are connected together if:

- the events share the same source IP address and/or User name,
- the target IP address of the first event is equal to the source IP address of the second,

The event correlation process would be realized during a training period. The resulting correlated events build an oriented graph, called users activity graph, where each node represents an event and each arc a causal relationship according to the correlation definition. Nevertheless, user's activity graph is too large and concentrated to be compliant with this technical constraint, we follow the recommendation defined in our work [33] by creating a "loop node" expressing recurrent events points in a sequence.

After the Bayesian Network structure creation, the training data set is used to compute conditional probabilities inside the Bayesian Network. We use the simple and fast counting-learning algorithm [37] to compute conditional probabilities considering each event as an experience.

The time duration between collected events can also be modeled by adding information in the Bayesian Network. The relation between clusters of events can be characterized by a temporal node holding the time duration probabilities between events. This extension is defined in detail in [35].

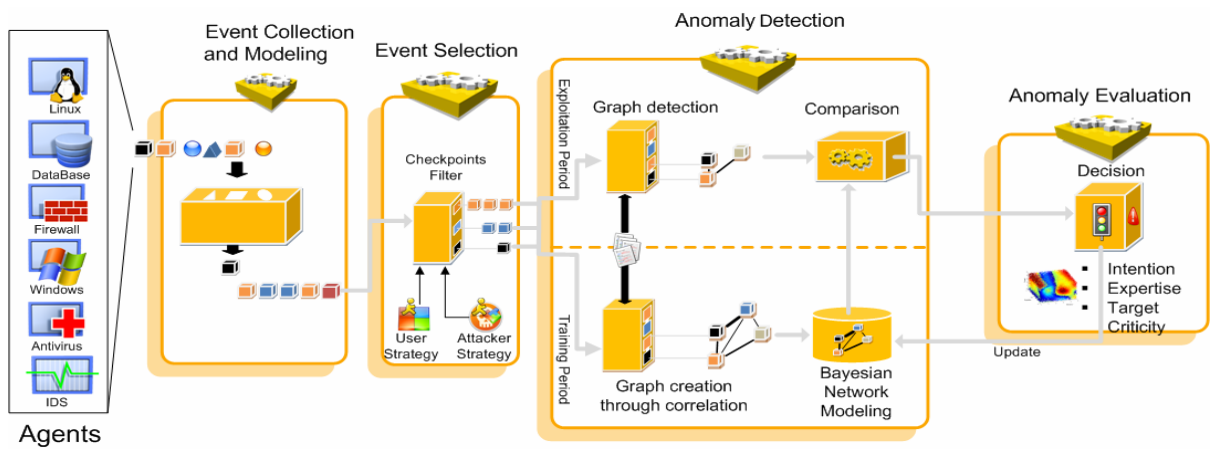

Fig. 4 Global Anomaly Intrusion Detection Architecture

efficient for a Bayesian structure. We propose a merging function that gathers events according to their normalization. Based on the edge contraction definition [10], we compute a new graph, called merged graph, where each node represents a cluster of events sharing the same meaning (same semantics) and holding a list of events attributes which identifies the owner of each event in this cluster. The resulting graph would show casual relationships between each user's events classes as described in section 5.1. This merged graph is used as the basis of our Bayesian structure.

Classical Bayesian Networks are built on acyclic oriented graph which is not the case here because of user activities periodical sequences of events. Although some methods exist to
allow Bayesian Network working on cyclic graph [23], most Bayesian libraries do not support cycles. To be

### 4.2 Anomaly Detection

The anomaly detection is performed during the exploitation period. This period intends to highlight abnormal behaviors between received data set and the trained user's activities model.

First of all, we compute a small user activities graph as defined in section 4.1.1 for a certain period of time represented by a temporal sliding windows on incoming events. This graph reflects all the users activities interactions for the sliding windows time period.

This detection graph is then compared with our normal user model (BN) which makes two types of deviances emerged: graph structure deviance detection and probabilistic deviance detection. To check the structure compliance between both graphs, we first control the correctness of each graph's features, and then we check if the detected event's classes belong to our reference, if the relationships between event's classes are valid and if the detected event's attributes are valid. Finally, each step of each sequence of events inside the detection graph is injected in the Bayesian Network. For each step, our model evaluates the probability to receive a specific event knowing it precedence events, if this probability is below a threshold the events is considered as deviant. When a deviance is detected, an alert is sent to the security analyst.

### 4.3 Event Selection

To prevent from a graph overloading, i.e. an expensive memory and CPU consumption, we simplify our model to provide only the relevant information about User Behaviors [34].

Indeed, with a huge volume of distinct events coming from the monitoring IS, the complexity of user activities increases greatly. On large IS, lots of interactions can appear between user actions. In the worst case, the resulting graph can be a heavy graph. So, to specify the relevance of collected events, we studied the interactions between legitimate user's behaviors and attacker's strategy [34].

We define a new way to reduce the model's complexity of user or system activities representation. We introduce the notion of necessary transit actions for an attacker to achieve these objectives: these actions are called Checkpoints. Checkpoints are based on different classes of attacker scenarios;
User to Root, Remote to Local, Denial of Service, Monitoring/Probe and System Access/Alter Data. We enrich this attacks classification with classes of malicious actions (Denial of Service, Virus, Trojan, Overflow, etc). For each scenario, we provide a list of Checkpoints which determine all the necessary legitimate activities needed by an attacker to reach such effects.
For instance, to achieve a User to Root effect an attacker chooses between six different variants of scenarios (Gain, Injection, Overflow, Bypass, Trojan, Virus). A checkpoint related to an Injection 1 is, for example, a command launch. We analyzed all the checkpoints of all the possible actions leading to one of these five effects.
We propose a selection of thirteen checkpoints representing different types of events involved in at least one of the five effects. These checkpoints reflect the basis of the information to detect attacker's activities. We also provide a description of the context to determine if all checkpoints need to be monitored regarding to the nature and the location of a component.

We extract the core information needed to detect misuses or attack scenarios. Thus, we do not focus our work on all the data involved in misuse or variant of attack scenarios but only on one piece of data reflecting the actions shared by the user's and attacker's behavior. We also study a couple of sequences of actions selection following an identical consideration.

Both checkpoints and sequences selections provide a significant model complexity reduction. Indeed, we manage a reduction of 24% of nodes and 60% of links. This selection slightly reduces the detection rate of unusual system use and reduces false positive of 10%.

### 4.4 Anomaly Evaluation

The lack of classical anomaly detection system is mainly due to a high false positive rate and poor information description about deviant events. The majority of these false positive comes from the evolution of the normal system behavior. Without a dynamic learning process, anomaly models become outdated and produce false alerts. The update mechanism has been enhanced by some previous works [14,29] which point out two main difficulties.
The first difficulty is the choice of the interval time between two update procedures [14]. On one hand, if the interval time is too large, some Information System evolutions may not be caught by the behavior detection engine. On the other hand, if it is too small, the system learns rapidly but loses gradual behavior changes. We do not focus our work especially on this issue but we assume that, by modeling users' activities behaviors, a day model updating is a good compromise between a global User behavior evolution and the time consumption led by such updates.
The second difficulty is the selection of appropriate events to update the reference in terms of event natures. Differentiating normal behavior evolution from suspicious deviation is impossible without additional information and context definition. To take efficient decisions, we need to characterize each event through specific criteria. These criteria should identify the objective of the end user behind the deviating events. We focus our work on this second issue and follow the approach in [40] that analyzes end users security behavior.

Our evaluation process evaluates a deviating event through a three dimensions evaluation of each deviating events: the intention behind the event, the technical expertise needed to achieve the event and the criticality of the targeted components by the event, each dimension characterizes a property of the deviating event.

---

[1] An injection consists in launching an operation through a started session or service.

All received events are associated with one of the three types of movements introduced in section 5.1: the intention of the configuration which defines beneficial moves trying to improve the system, intention of activity which represents usual activity on the IS or neutral activity and then the intention of attack which refers to all the malicious activities in the system. The degree of deviation of an event would inform us how far an event from the normal use of the system is. We assume that the more an event is far from normal behavior, the more this deviating event holds malicious intention. Finally, other past events linked by casual relationship with the deviating one lead also the malicious intention. The expertise dimension defines the technical expertise needed by a user to realize an event. This expertise is computed on the type of actions realized by the event (action of configuration or action of activity), the type of a targeted component (a Router needs more technical expertise than a Work Station) and the owner of the event (classical user or administrator).

Finally, the event's impact on IS will depend on the targeted component. Thus, we evaluate a deviating event also by the criticality of the targeted component. This criticality is evaluated by combining vulnerabilities held by the targeted component, the location of the targeted component (e.g. LAN, Public DeMilitary Zone, etc) and its business importance (e.g. critical authentication servers are more important than workstations regarding the business of the company).

According to all dimensions definitions, each deviating point will be located in this three dimension graph. The three dimension representation increases the analyst visibility of deviating events. Nevertheless, some automatic actions could considerably help analyst to decide if a deviant event is related to a normal system evolution or to intrusive activity. We propose to build a semi-automatic decision module to define which deviating events fit normal system evolutions and which ones reflect attackers' activities. Our semi-automatic decision module is a supervised machine learning method. We used a training data set composed of deviating events located in our three-dimension graph. Each point of the training data set is identified as a normal system evolution or attackers' activities by security analysts. To learn this expertise knowledge, we use a Support Vector Machine (SVM) to define frontiers between identified normal points and attack points. The SVM technique is a well known classification algorithm [20] which tries to find a separator between communities in upper dimensions. SVM also maximizes the generalization process (ability to not fall in an over-training). After the construction of frontiers, each deviating events belonging to one community will be qualified as a normal evolution or attackers' activity and will receive a degree of belonging to each community. After that, two thresholds will be computed. They define three areas; normal evolution of system area, suspicious events area and attack or intrusive activities area. Each deviating events belonging to normal evolution area will update our normal Bayesian model. Each deviating events belonging to the intrusive activities area or suspicious area will create an alarm and send it to an analyst.

## 5. Experimentation

In this section, we aim to make into practice the two proposed modules, event modeling and User Behavioral Analysis, while using a large corpus of real data. Event modeling experience will normalize raw events in the ontology's categories that describe her semantics. User behavioral analysis experimentation will use normalized events generated by events modeling module to detect abnormal behaviors.

### 5.1 Event Modeling

To study the effectiveness of the modeling proposed in Section 5.1, we focused our analysis on the exhaustiveness of the ontology (each event is covered by a category) and on the reduction of event number to be presented to the security analyst. We performed an experiment on a corpus of 20182 messages collected from 123 different products.

| Security logs(51) | Applications logs(29) | System logs(22) | Network device logs(7) |
|---|---|---|---|
| IDS(5) Firewall(18) Antivirus(9) Authentication(8) VPN(3) IPS(2) Antispam(1) Monitoring(5) | Mail(7) FTP(3) Web(3) Database(2) DNS(1) Proxy(7) Others(6) | Windows(4) Unix-Linux(14) Autre(4) | Router(1) Switch(1) Management Console(5) |

Fig 5: Type of used product

The main characteristic of this corpus is that the events are collected from heterogeneous products, where the products manipulate different concepts (such as attacks detection, virus detection, flaws filtering, etc.). The sources used are security equipment logs, audit system logs, audit application logs and network component logs. Figure 5 illustrates the various probes types used and, into brackets, the number of probes per type is specified. The classification process was performed manually by the experts. The expert reads the message and assigns it to the category which describes its semantics. The expert must extract the intention from the action which generated the message, the movement used to achieve the intention, the target toward which the intention was directed and the gain related to this intention.

We have obtained, with the manual classification of raw events, categories of various sizes. The distribution of the messages on the categories is represented on figure 6. Some categories are large, the largest one contains 6627 events which presents a rate of 32,83% of the corpus. This is due to the monitoring of the same activity by many products or to the presence of these signatures in many products.

supervised by these products. The conclusion that we can draw from this study is that a good line of defense must supervise all the activities aimed in an IS, and that the cooperative detection should not be focused on the number of the deployed products but on the activities to be supervised in the IS. This result can bring into question the choice of the defense line for the IS.

5.2 Behavioral Analysis

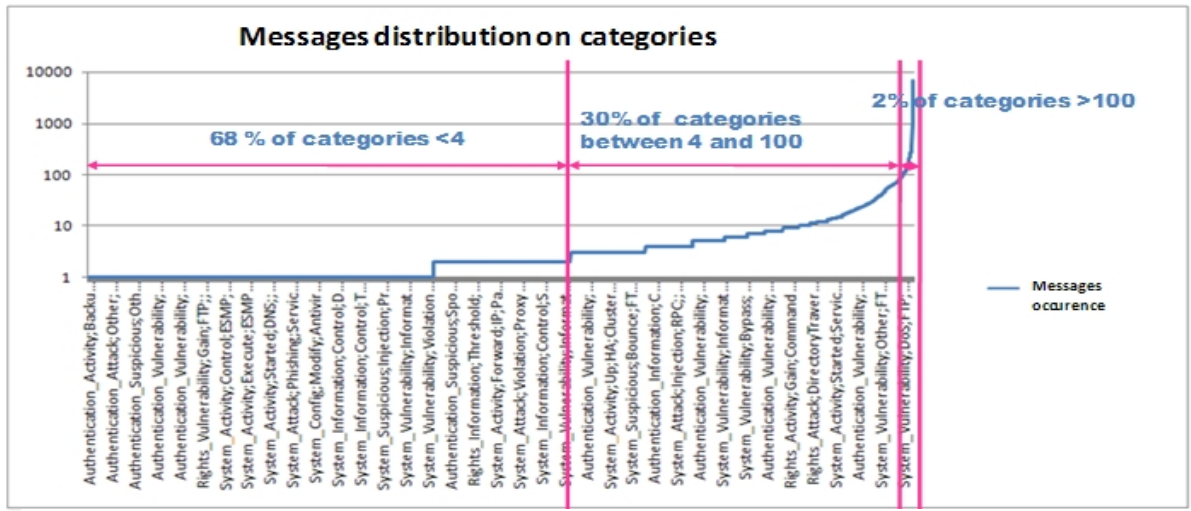

Fig. 6 Events Distribution on the Ontology's Categories

The representation of the events under the same semantics reinforces the process of managing the security in a cooperative context and facilitates the task of the analyst (more detail in [2]). In addition, we had a singleton categories, 732 raw events forming their own category, which represent a rate of 42,21% of all categories and which represent only 3,63% of the corpus. Event modeling has reduced the number of events by 91,40% (from 20182 to 1734). The presence of singleton categories can be explained by the following points: only one product among the deployed products produces this type of event. A signature, which is recognized by a product and not recognized by an another, errors made by experts generated the creation of new categories, they do not have to exist theoretically, and the presence of targets monitored rarely increases the number of singleton categories, because the movement exists several times, but only once for these rare targets.

We observe that the category of the movement Suspicious introduced into our ontology is quite necessary to preserve the semantics of a raw event which reflects a suspicion. These types of events will be processed with the User Behavioral Analysis. Ontology does not make it possible to analyze event, its goal is to preserve raw events semantics. The proportions of the various categories depend on the deployed products and the activities to be

Actual Intrusion Detection System operating on a Global Information System Monitoring lacks of large test dataset aiming at checking their efficiency and their scalability. In this section, we provide our results on our Anomaly Intrusion Detection System using a real normalized data set. We deployed our architecture on a real network and collected events coming from hundreds of IS components.

**DataSet Analysis:** Our dataset analysis comes from a large company composed of hundreds of users using multiple services. The dataset has been divided into two datasets: one training data set composed of events collected for 23 days and the other one (test data set) composed of events collected for 2 days after the training period. The training dataset aims at train-ing our engine and creating a user normal behavioral model. The test 'dataset' has been enriched of attack scenarios in or-der to test our detection engine. First of all, the test data set is used to test the false alarms rate (false positive rate) of our engine. Then attack scenarios will be used to determine our detection rate, and more over the false negative rate (not detected attacks rate) of our engine.

The major parts of the collected events are web server information and authentication information. We can notice that during the monitored period, some types of events are periodic (like Authentication_Activity .Login.

SysAuth.Account.Success) and other ones are sporadic. Moreover, our dataset is composed of more than 70 types of events ranging from Authentication actions to System usage (like service start).

Our training dataset, representing the activity of the system for 23 days, is composed of 7 500 000 events. The test data is composed of 85 000 normal events (two days of the System's activity) and 410 events representing three different attack scenarios. These scenarios reflected three types of attack effects on the system as introduced in the DARPA attacks classification (Remote to local, User to Root,...). Some scenario variants are developed for each class. For example, concerning the Remote to Local attack scenario, we provide two kinds of variant of scenario as follow:

The remote to local variant one is composed of four different classes of events:
-Authentication_Activity.Login.SysAuth.Account.Success,
-uthentication_Config.Add.SysAuth.Account.Success,
-Authentication_Activity.Login.SSH.Admin.Success,

different times. Nodes (referring to event's classes) become stationary around the step 330 whereas links (relationships between event's classes) continue to evolve until step 360. Only the status (user or process identifier) seems to never reach a stationary point. To understand this phenomenon, we analyze in depth the evolution of the status of each different nodes. We notice that the status of one particular node, Authentication_Activity .Login.SysAuth.Account.Success, blow up. We investigate and discover that the considered company owns an e commerce Web server on which each new consumer receives a new login account. That is why when other nodes reach their stationary point around the 390th step, Authentication_Activity. Login. SysAuth.Account. Success node continues to grow. To avoid a complexity explosion inside our Bayesian model, we add a constraint defining a time of unused events indicator. We define a threshold to determine which state of node will be kept and which one will be dropped.

| Scenarios | Type | Events number | Deviance of node | Deviance of state | Probabilistic deviance | | | | | | | | | | |
|---|---|---|---|---|---|---|---|---|---|---|---|---|---|---|---|
| | | | | | 0,1 | 0,01 | 0,001 | 0,0001 | 0,002 | 0,1000 | 0,0100 | 0,0010 | 0,0001 | 0,0020 | |
| Normal Events | - | 85000 | | | 68688 | 14757 | 4556 | 394 | 11907 | 0,8081 | 0,1736 | 0,0536 | 0,0046 | 0,1401 | False positive Detection rate |
| User to Root | sc1 | 240 | 110 | 118 | 10 | 10 | 2 | 0 | 10 | 0,9917 | 0,9917 | 0,9583 | 0,9500 | 0,9917 | |
| | sc2 | 40 | 0 | 17 | 11 | 11 | 0 | 0 | 8 | 0,7000 | 0,7000 | 0,4250 | 0,4250 | 0,6250 | |
| | sc3 | 30 | 10 | 8 | 9 | 9 | 1 | 1 | 9 | 0,9000 | 0,9000 | 0,6333 | 0,6333 | 0,9000 | |
| Remote to Local | sc1 | 20 | 0 | 11 | 7 | 7 | 2 | 1 | 6 | 0,9000 | 0,9000 | 0,6500 | 0,6000 | 0,8500 | |
| | sc2 | 30 | 0 | 18 | 13 | 13 | 6 | 4 | 11 | 1,0333 | 1,0333 | 0,8000 | 0,7333 | 0,9667 | |
| Access/alter data | sc1 | 20 | 0 | 9 | 0 | 0 | 0 | 0 | 0 | 0,4500 | 0,4500 | 0,4500 | 0,4500 | 0,4500 | |
| | sc2 | 30 | 0 | 16 | 11 | 11 | 2 | 0 | 8 | 0,9000 | 0,9000 | 0,6000 | 0,5333 | 0,8000 | |
| | | | | | | | | | | 0,9220 | 0,9220 | 0,8049 | 0,7878 | 0,9000 | Attack Detection rate |

Fig. 7 Detection Sums

-System_Activity.Stop.Audit.N.Success.
The second one holds the classes below:
-Authentication_Activity.Login.SysAuth.Account.Success,
-Authentication_Config.Modify.SysAuth.Account.Success,
-Authentication_Activity.Login.SysAuth.Admin.Success,
-System_Activity.Execute.Command.Admin.Success.

Each variant of each scenario is reproduced ten times with different attributes (login user, IP address Source and Destination) belonging to the data set. We can notice that all events involved in attack scenarios refer to legitimate actions. All these events define a set of event among shared actions between legitimate user behaviors and attacker strategy.

**Results:** The test data set is used to build our user activities model. To compute efficiently this model, we split the training data set into 440 steps.

Each Bayesian structure feature (nodes, links, states) evolves differently and reaches its stationary point at

The test data set is then processed by our Anomaly detection System and our detection's results are in figure 7. This table distinguishes each scenario's events and the detection rate for different probability threshold. These thresholds could be chosen regarding to the organisms or company goals. In case of a very sensitive IS, the attack detection rate needs to be as high as possible. A probability threshold of 0.002 achieving a detection rate of 90% with false positive rate around 14% would be suitable. In case of a more transversal use of our approach, companies deal with false positives and detection rate. A threshold of 0.0001 provides an attack detection rate of 79% with a false positive rate below 0.5%. Additional observation can be made regarding our attack detection rate. Most of the time, attack detection rate of detection tools reaches 95% but in our context, all our scenarios are composed of events which belong to normal behavior. All these events do not necessary deviate from the normal

IJCSI

behavior that is why our detection rate is slightly below classical detection rate. We can estimate that little less than 10% of the test attack's events belong to normal behavior (legitimate event and attributes). Despite this constraint, we still reach detection rate from 80% to 90%.

## 6. Conclusion and Perspectives

Our main goal throughout this paper was to describe a framework that addresses companies: managing the company's security information to protect the IS from threats. We proposed an architecture which provides a global view of what occurs in the IS. It is composed of different agent types collecting and modeling information coming from various security products (such as firewalls, IDS and antivirus), Operating Systems, applications, databases and other elementary information relevant for the security analysis. An Analysis Server gathers all information sent by agents and provides a behavioral Analysis of the user activities.

A new modeling for describing security information semantics is defined to address the heterogeneity problem. The modeling is an ontology that describes all activities that can be undertaken in an IS. By using real data triggered from products deployed to protect the assets of an organization, we shown that the modeling reduced the amount of events and allowed automatic treatments of security information by the analysis algorithms. The model is extensible, we can increase the vocabulary according to the need such as adding a new movement to be supervised in the IS. The model can be applied to other contexts of monitoring such as the monitoring of physical intruders in a museum; all we have to do is to define the adequate vocabulary of the new context.

We demonstrated that unknown attack scenarios could be efficiently detected without hard pre description information through our User Behavioral Analysis. By using only relevant information, User's behaviors are modeled through a Bayesian network. The Bayesian Network modeling allows a great detection effectiveness by injecting incoming events inside the model and computing all conditional probabilities associated. Our Anomaly evaluation module allows updating dynamically a User's model, reducing false positive and enriching Behavioral Anomalies. The experimentation on real data set highlights our high detection rate on legitimate action involved in Attack scenarios. As data are modeled in the same way, User Behavioral Analysis results show that the effectiveness of the analysis processes is highly dependent on the data modeling.

The proposed framework can be useful to other processes. Indeed, the ontology is necessary to carry out counter-measures process, the results of User Behavioral Analysis allowing the administrator to detect legitimate users that deviate from its behavior, a reaction process can then be set up to answer malicious behaviors.

**Jacques Saraydaryan** holds a Master's Degree in Telecoms and Networks from National Institute of Applied Sciences (INSA), Lyon –France in 2005, and a Ph.D. in computer sciences from INSA, Lyon France in 2009. He is a Research Engineer at the Exaprotect company, France. His research focus is on IS Security especially on Anomaly intrusion detection system. His research work has been published in international conferences such as Secrureware'08, Securware'07. He has one patent with Exaprotect Company.

**Fatiha Benali** holds a Master's Degree in Fundamental Computer Sciences at Ecole Normale Supérieure (ENS), Lyon -France, and a Ph.D. in computer sciences from INSA, Lyon-France in 2009. She is a Lecturer in the Department of Telecommunications Services & Usages and a researcher in the Center for Innovations in Telecommunication and Services integration (CITI Lab) at INSA, Lyon- France. Her research focus on IS security notably on information security modeling. Her research work has been published in international conferences such as Security and Management (SAM'07), Securware'08. She has 2 papers awarded and one patent with Exaprotect Company.

**Stéphane Ubéda** holds a PhD in computer sciences at ENS Lyon - France in 1993. He became an associated professor in the Jean-Monnet University-France in 1995, obtain an Habilitation to conduct research in 1997 and became in 2000 full professor at the INSA of Lyon. He is a full professor at INSA of Lyon in the Telecommunications department. He is the director of the CITI Lab, he is also the head of the French National Institute for Research in Computer Science and Control (INRIA) Project named AMAZONES for AMbient Architectures: Service-Oriented, Networked, Efficient, Secure.